# Fundamentals of PV Efficiency: Limits for Light Absorption

M. Ryyan Khan, Xufeng Wang, and Muhammad A. Alam

*Abstract*—A simple thermodynamic argument related to a (weakly absorbing) finite dielectric slab illuminated by sunlight – originally suggested by Yablonovich – leads to the conclusion that the absorption in a dielectric can at best be increased by a factor $4n^2$. Therefore, the absorption in these materials is always imperfect; the Shockley-Queisser limit can be achieved only asymptotically. In this paper, we make the connection between the degradation in efficiency and the Yablonovich limit explicit and re-derive the $4n^2$-limit by intuitive geometrical arguments based on Snell's law and elementary rules of probability. Remarkably, the rederivation suggests strategies of breaking the traditional limit and improving PV efficiency by enhanced light absorption.

*Index Terms*— Photovoltaic cells, thermodynamics, intensity, absorption.

## I. Introduction

THE thermodynamic argument proposed by Shockley-Queisser (S-Q) [1] allows us to calculate the maximum efficiency of solar cells. In an earlier paper [2], we have shown that the essential features of the thermodynamic limit (as well as various practical approaches proposed to approach or exceed it) can be understood by shining sunlight onto a box of atoms characterized by two energy levels, $E_1$ and $E_2$, see Fig. 1.

At equilibrium, a system comprising of a collection of 2-level atoms can be described by the Fermi-Dirac (F-D) statistics:

$$f_i = \frac{1}{e^{\theta_i/T_D}+1}.$$

Here $k_B\theta_i = E_i - \mu_i$ with $E_i$ and $\mu_i$ being the energy and chemical potential of the $i$-th state $(i=1,2)$. On the other hand, the photons follow Bose-Einstein (B-E) distribution given as follows,

$$n_{ph}(T) = \frac{1}{e^{[(E_1-E_2)-\Delta\mu]/k_BT}-1}.$$

For the sun, $\Delta\mu \approx 0$. The open circuit voltage of the solar cell made out of the system of atoms is given by the splitting of the chemical potentials, i.e., $qV_{OC} = (\mu_1 - \mu_2)$.

Authors are with the Electrical and Computer Engineering Department, Purdue University, West Lafayette, IN 47906 USA (e-mail: alam@purdue.edu).

The S-Q limit [1] can be theoretically achieved only if all the photons of right energy $(\hbar\omega = E_1 - E_2)$ entering the dielectric are absorbed with probability one. Our previous derivation of the S-Q limit for the 2-level system presumed perfect absorption [2]. For imperfect absorption, we should have written the upward and downward transition rates as,

$$U = P \times \theta_S \; f_2(1-f_1)n_{ph} \qquad (1)$$

$$D = 1 \times \theta_D \; f_1(1-f_2)(n_{ph}+1) \qquad (2)$$

allowing for the fact that some photons of the right energy may exit the solar cell without being absorbed $(P<1)$. Here, $n_{ph}$ is the B-E distributions related to the radiation from the sun (with appropriate $\Delta\mu$ and $T$). $\theta_S$ and $\theta_D$ are the input and output radiation angles. If we equate Eq. (1) and (2) and follow the procedure in Ref. [2] to recalculate the efficiency $(\eta)$ of the simplified 2-level model, we find

$$\eta = \left(1 - \frac{T_D}{T_S}\right) - \frac{kT_D}{E_g}\log\left(\frac{\theta_D}{\theta_S}\right) - \frac{kT_D}{E_g}\log\left(\frac{1}{P}\right). \qquad (3)$$

A reduction in absorption $(P)$ reduces $\eta$ below the thermodynamic limit, an intuitive result.

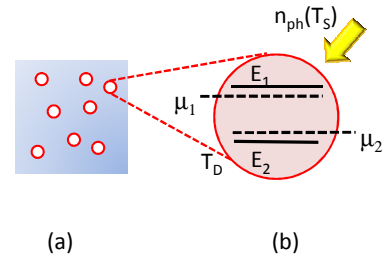

Figure 1: (a) A system comprising of 2-level atoms, (b) A 2-level atom interacting with incident light.

In this paper, we will show that absorption probability (absorptance) for the weakly absorbing material is given by

$$P = \frac{f_A \times \alpha L}{(f_A \times \alpha L)+1}, \qquad (4)$$

where $L$ is the thickness of the dielectric slab defined by the electrical design of the cell, $\alpha$ is the absorption coefficient the material under consideration, and $f_A$ is the absorption enhancement factor defined by optical design of the cell. For weakly absorbing $(\alpha \to 0)$ material such as silicon (near band-edge), $f_A$ and/or $L$ must be enhanced somehow to make $P \rightarrow 1$ and the efficiency approaches the 2-level efficiency limit (analogous to S-Q limit), see Eq. (3). Unfortunately, the cell cannot be made arbitrarily thick, because the photo-generated carriers in a thick film will recombine before being collected by the contact, and the short-circuit current $J_{SC}$ will be reduced. Instead, we should focus on increasing $P$ by increasing $f_A$, with clever arrangement of mirrors, reflectors, concentrators, photonic crystals and metamaterials.

In 1982, Yablonovich used the theory of detailed balance of photons to provide a surprising answer [3][4]: In essence, no matter how clever or sophisticated the optical design, $f_A$ cannot exceed $4n^2$ ($n$ is the refractive index of the solar cell), and therefore, the absorption probability $P$ of a finite cell of thickness $L$ can never be perfect. The theory suggests several practical ways to approach the limit. For a poorly designed cell, $f_A \rightarrow 1$, $f_A \alpha L \ll 1$, so that $P \sim \alpha L$; a more thoughtful design enhances $f_A \rightarrow 4n^2$, so that even with $f_A \alpha L < 1$ $P \sim 4n^2 \alpha L$. While $P < 1$, but at least a good design could increase $P$ by a factor of $4n^2$ over a poorly designed one.

In Sec. II, we re-derive Yablonovich limit by elementary geometrical arguments based on Snell's law. We explain the absorption enhancement factor $f_A (= f_I \times f_L)$. Here, $f_L$ is the average absorption path length enhancement factor per trip through the dielectric (normalized to cell thickness) to account for the fact that rays at an random angle $\theta_i$ has a higher probability of absorption compared to a ray passing vertically through the cell, i.e., $f_L \equiv \langle L_i^{eff}(\theta_i) \rangle / L$. And, $f_I = 2\beta$ is the intensity enhancement factor calculated from the average number of bounces $\beta$ a photon experiences before it escapes the cell. In turn, $\beta$ depends on material index $n$ and the dimensionality of surfaces $D_S$ defining scattering of photons. Therefore, $f_A$ in a weakly absorbing material is determined by determining by simple geometrical arguments the two parameters, $\langle L_i^{eff}(\theta_i) \rangle$ and $\beta$. The derivation of Eq. (4) will also suggest techniques to beat the Yablonovich limit, i.e., $f_A > 4n^2$, by restricting the emission angle or changing the statistics of photon scattering, see Sec. III. Some of these have also been discussed in [5], however, we offer intuitive interpretations and significant generalizations of the key results to conventional structures and some recent light trapping configurations. Our conclusions are summarized in Sec. IV.

## II. STATISTICS OF LIGHT RAYS

### A. A Summary of the key Results

The essence of the Yablonovich's argument in [3][4] is understood by the following scenario: Consider a dielectric with an inlet and an outlet for photons. At steady state, the incoming and outgoing flow rates (#/sec) are equal to ensure there is no constant buildup of energy inside the dielectric. If the incoming flux (#/sec/area), $F_i$ is fixed, the steady state number of photons inside the layer can be controlled by changing the outgoing flux, $F_o$. In principle, the arbitrary decrease in $F_o$ would be assisted by a corresponding increase in the photon density inside the dielectric layer. Generalizing Yablonovich's argument, we can show that when a solar cell is illuminated by sunlight of flux $F_i$ (Fig. 2), the emission (outgoing flux) cannot be reduced arbitrarily by optical design and therefore, the 'pressure' of photon gas inside the cell (i.e., intensity enhancement) cannot exceed

$$f_I \sim \hat{C}_{DS} \times n^{D_S} \times \left(\frac{\theta_C}{\theta_{esc}}\right)^{D_S}.$$

This limit is dictated solely by the refractive index $n$ of the dielectric, dimensionality $D_S$ of the photon scattering surface and the maximum escape angle, $\theta_{esc}$. Here, $\theta_C$ is the critical angle for total internal reflection. For typical dielectrics, $\theta_{esc} = \theta_C$, but angle-selective photonic crystal can reduce $\theta_{esc}$ below $\theta_C$ to enhance the intensity within the dielectric, as explained in Sec. III. Here, $\hat{C}_{DS}$ is a dimensionality constant.

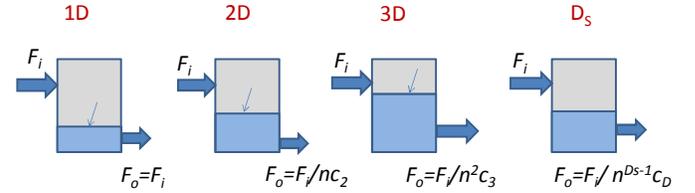

**Figure 2: Input and extraction of photons to fill up a 'container/photon-tank'.**

For a weak absorbing dielectric, the absorption only occurs as a small perturbation to the intensity and $f_L$. Hence the absorption enhancement differs from the intensity enhancement by only a factor:

$$f_A(n, D_S) \sim C_{DS} \times n^{D_S} \times \left(\frac{\theta_C}{\theta_{esc}}\right)^{D_S}. \quad (5)$$

As we will see below that in a conventional 3D solar cells with a Lambertian back mirror, $D_S = 2$, $C_{DS} = 4$ and $\theta_{esc} = \theta_C$, and hence $f_A(n, D_S = 2) \sim 4n^2$. Eq. (5) can be inserted in Eq. (4) to calculate absorptance for different types of optical configurations.

### B. Derivation of Eq.(4) for Classical Cells

Consider the fate of a photon trapped within a finite dielectric slab as it tries to escape the dielectric region by repeated bouncing between the two (one reflecting, one random) surfaces, as in Fig. 3(a). Snell's law ($n_1 \sin\theta_1 = n_2 \sin\theta_2$) dictates that the maximum angle with which a photon incident onto dielectric/air interface can escape the dielectric is given by $\theta_C = \sin^{-1}(1/n)$. The probability that a ray will escape ($P_{esc}$) through the escape cone ($0 < \theta < \theta_C$) depends on the dimensionality of the confining surfaces (Fig. 3).

If a ray is incident outside the escape cone, it will bounce back, for total internal reflection, in a random angle defined by the local orientation of the top interface. The average number of bounces a photon experiences before it escapes the dielectric is defined by the escape probability *per bounce* as

$$\beta = P_{esc}^{-1}. \quad (6)$$

Note that the number of bounces before escape equals the enhancement of photon intensity per round trip inside the dielectric layer. Figure 4 explains why: if a photon bounces $\beta$ times before it escapes, any arbitrary point 'A' within the dielectric is visited by $\beta$ (on average) number of photons, so that the intensity of the point goes up by the same factor. Therefore, $f_I = 2\beta$ for a cell with a mirror in the back; or, $f_I = \beta$, if the mirror is absent.

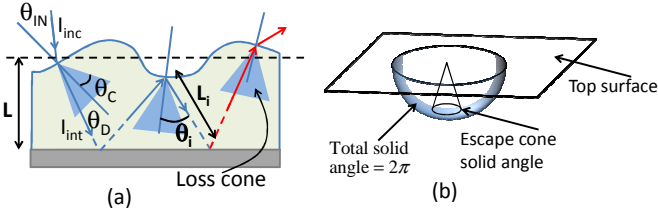

**Figure 3:** (a) Definition of loss-cone (escape cone) and path lengths, (b) Illustration of escape cone solid angle in a 3D case.

It should be understood that intensity enhancement ($f_I = 2\beta$) does not arise from an enhanced emission rate from atoms within the cell. The emission rate of photons depends only on the temperature of the dielectric (blackbody radiation). By reducing the probability of escape ($P_{esc}$) with suitably designed structures, the number of photons inside ($N_{ph}$) is increased, as in Fig. 2, so that the number of photons escaping the cell ($P_{esc} \times N_{ph}$) remains independent of the escape angle.

Within a given bounce, the rays are scattered at different angles ($0 < \theta_i < \pi/2$), see Fig. 3(a). On average, the path length per pass (up or down), $\langle L_i^{eff}(\theta_i) \rangle$, is greater than $L$, the thickness of the cell. Therefore, the absorption per round trip, with a back mirror, is $\alpha \times 2\langle L_i^{eff} \rangle \equiv 2f_L \alpha L$ where $f_L \equiv \langle L_i^{eff} \rangle / L$.

Now to complete the derivation of Eq. (4), consider a solar cell in which photons bounce $\beta$ times between top and bottom interfaces before exiting a dielectric slab of length $L$ and absorption coefficient $\alpha$. The probability of absorption per round trip is $\sim 2f_L \alpha L$ and in every round-trip, a fraction $1/\beta$ of the photons escape through the top surface without being absorbed. Therefore, the absorption probability or absorptance is,

$$P = A = \frac{2f_L \alpha L}{2f_L \alpha L + \beta^{-1}}$$
$$= \frac{2f_L \beta \alpha L}{2f_L \beta \alpha L + 1} \equiv \frac{f_L f_I \alpha L}{(f_L f_I \alpha L) + 1} \equiv \frac{f_A \times \alpha L}{(f_A \times \alpha L) + 1}.$$

Once the two parameters, $\beta(= P_{esc}^{-1})$ and $f_L(L^{eff})$, are determined for a given optical configuration, $\eta$ (of the 2-level solar cell system) is readily determined through Eqs. (3) and (4). We now calculate $\beta$ and $f_L(L^{eff})$ for a few typical optical configurations.

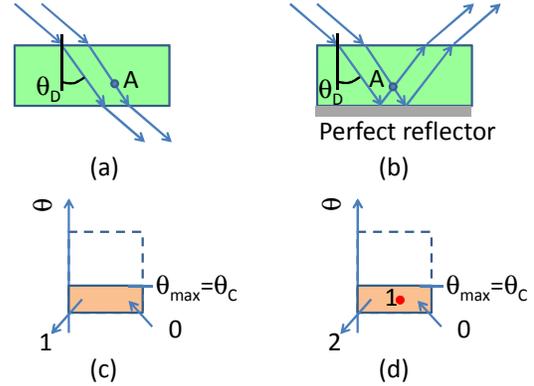

**Figure 4:** Angle statistics (c, d) of photons in a 1D object (a) without and (b) with a back reflector.

### C. A Planar Bottom Mirror with no Randomness ($D_S=0$)

Consider a dielectric defined by two parallel, planar surfaces. If a ray of sunlight refracts into the dielectric as in Fig. 4(a), it must enter the dielectric within the 'escape cone' (see arrow labeled '0' in Fig. 4(c)). If the ray is neither decayed (negligible perturbation due to absorption), nor scattered within the dielectric, the ray will be incident on the bottom surface within the escape cone and will escape to air (arrow '1', Fig. 4(c)), with no further reflection. The path length enhancement depends on the angle of the refracted ray ($\theta_D$) inside the dielectric, i.e., $f_L = 1/\cos\theta_D$. The number of trips through the dielectric is just 1 ($\beta = 1$). Only a single ray is associated with any point 'A', so that the density of photons at each point inside the dielectric is exactly equal to that in air ($f_I = \beta = 1$). Therefore, the absorption enhancement is $f_A = f_I f_L = \beta f_L = 1/\cos\theta_D$ —an intuitive result. Note that if the photon is never absorbed or scattered within the dielectric, its angle cannot change, and the dashed region in Fig. 4(c) remains forever inaccessible. For practical dielectrics $\theta_D (<\theta_C) \to 0$ and $\cos\theta_D \to 1$.

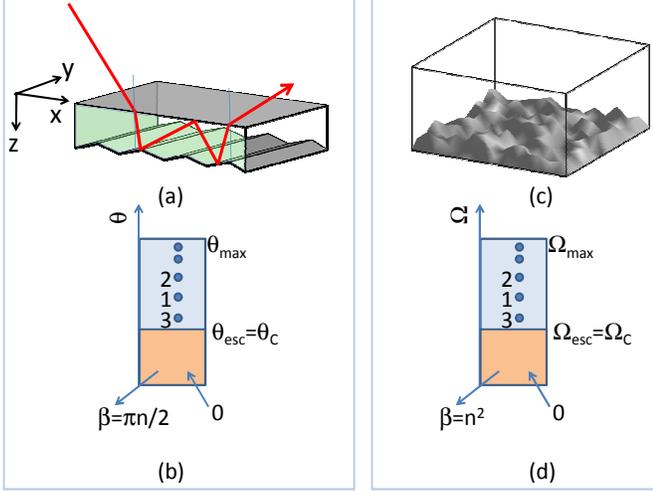

**Figure 5: Angle statistics for photon scattering in (a) 1D and (c) 2D Lambertian reflector at the back. The corresponding angle statistics are shown in (b) and (d).**

If we make the back-surface fully reflecting, as in Fig. 4(b), the ray still enters the dielectric within $\theta_{max} = \theta_C$ (Fig. 4(d)), bounces once on the back mirror(red dot, marked '1'), and then escapes through the top interface (arrow marked '2') – never once leaving the escape cone. The photon makes two trips ($f_I = 2$) through the dielectric before it escapes, so that $f_A = 2/\cos\theta_D = 2\beta f_L = f_I f_L$. By blocking off the exit from the bottom, the internal photon density has been raised by a factor of 2, because every point 'A' is traversed by two rays—one on its way down to the mirror, the other after bouncing back, on its way to escape.

The results above are consistent with Eq. (5), with $C_{DS} = 1$ (without mirror) and $C_{DS} = 2$ (with mirror). Since $D_S = 0$ for both cases, Eq. (5) suggests that $f_A$ is independent of the index of the dielectric, consistent with the results derived in the preceding paragraph. Finally, using (4) we find

$$P = A = \frac{2 f_L \alpha L}{2 f_L \alpha L + \beta^{-1}} = \frac{2\alpha L / \cos\theta_D}{2\alpha L / \cos\theta_D + 1} \approx \frac{2\alpha L}{2\alpha L + 1}.$$

Because in the absence of any scattering, $f_L = 1/\cos\theta_D \approx 1$. It is easy to see that a dielectrics defined by parallel, planar surfaces make a poor absorber, i.e., $P \sim 2\alpha L$ when $\alpha L$ is small. Absorption is enhanced considerably by roughening the bottom reflector, as discussed below.

### D. Bottom Mirror with 1D Randomness ($D_S$=1)

Fig. 5(a) shows a dielectric with a planar surface on the top and a roughened (only along the x direction), fully reflective surface in the back. Let us assume that the incident ray is restricted to planes parallel to xz-plane. The incident ray enters into the dielectric through the top planar surface within the escape cone ($\theta \leq \theta_C$) of the top surface. This is represented as state-0 in Fig. 5(b). The ray is scattered and reflected by the bottom rough surface. If the angle following the scattering is outside the escape cone ($\theta > \theta_C$), the state of the ray is characterized by a point within the blue region in Fig. 5(b). Since the ray is outside the escape cone, it will be internally reflected from the top surface (total internal reflection). The bouncing between the surfaces will continue and the photon will remain trapped within the dielectric, as long as the ray occupies a state outside the escape cone (the blue region in Fig. 5(b)). Statistically, on average, the photon will bounce $\beta$ times (i.e., hop through $\beta$ number of states in Fig. 5(b)) before it is randomly scattered into the escape cone and finally exits the structure (arrow-$\beta$ as shown in Fig. 5(b)). Note that $\beta$ should be understood as an average, because some photons may escape only a single bounce, while others may be trapped for bounces much larger than $\beta$.

The escape probability and $\beta$ can be calculated as follows:

$$P_{esc} = \frac{\int_{-\theta_C}^{\theta_C} \cos\theta \, d\theta}{\int_{-\pi/2}^{\pi/2} \cos\theta \, d\theta} = \frac{2\sin\theta_C}{2} = \frac{1}{n}$$

$$\beta = \frac{1}{P_{esc}} = n.$$

The intensity enhancement is $f_I = 2\beta = 2n$.

For a very weak absorbing dielectric, the absorption in a single pass by the randomly scattered light is

$$A_{single-pass} = \frac{\int \left(1 - e^{(-\alpha L/\cos\theta)}\right) \cos\theta \, d\theta}{\int \cos\theta \, d\theta}$$

$$\approx \frac{\int_{-\pi/2}^{\pi/2} \alpha L \, d\theta}{\int_{-\pi/2}^{\pi/2} \cos\theta \, d\theta} = \frac{\pi}{2} \alpha L.$$

Therefore the absorption path enhancement is

$$f_L = \frac{A_{single-pass}}{\alpha L} = \frac{\pi}{2}.$$

Also, the absorptrion enhancement factor is $f_A = f_I f_L = \pi n = C_1 n$, again consistent with Eq. (5). Of course, we still have $\theta_{esc} = \theta_C$.

Therefore, by Eq. (4) we find

$$P = A = \frac{2 f_L \alpha L}{2 f_L \alpha L + \beta^{-1}} = \frac{\pi n \alpha L}{\pi n \alpha L + 1} \quad (7)$$

for reflective rough surface (in one direction). The appearance of $\pi$ and $n(>1)$ suggests improved absorption – even for a poor absorber like silicon $(n = 3.49)$, the 1D rough bottom reflector increases absorptance by a factor of 11. And a $200 \mu m$ thick silicon layer will appear optically as a 2mm thick film.

### E. Random Surfaces ($D_S=2$)

It is possible to roughen the bottom reflector in both the *x* and *y* directions, see Fig. 5(c). The light comes in through the top planar surface and gets scattered by rough back reflector. The scattering of light and the trapping concept is the same as explained in the previous section. The light cannot escape from the dielectric if it is scattered into and then stays within the states in the blue region of Fig. 5(d). The light escapes after $\beta$ bounces, when a random scattering by the bottom interface scatters the ray into the escape cone.

To calculate $P_{esc}$ and $\beta$, we will integrate over the solid angle (3D) with $\theta_{esc} = \theta_C$, as follows:

$$P_{esc} = \frac{\int \cos\theta \, d\Omega}{\int_{half-sphere} \cos\theta \, d\Omega}$$
$$= \frac{\int_0^{2\pi} \int_0^{\theta_C} \cos\theta \sin\theta \, d\theta \, d\phi}{\int_0^{2\pi} \int_0^{\pi/2} \cos\theta \sin\theta \, d\theta \, d\phi}$$
$$= \sin^2 \theta_C \equiv 1/n^2.$$
(8)

So that,
$$\beta = \frac{1}{P_{esc}} = n^2.$$

And, the corresponding intensity enhancement is $f_I = 2\beta = 2n^2$. In silicon, on average, the ray travels an astonishing ~25 times inside the layer before it can escape.

For a very weak absorbing dielectric, the absorption in a single pass by the randomly scattered light is

$$A_{single-pass} = \frac{\int (1 - e^{(-\alpha L/\cos\theta)}) \cos\theta \, d\Omega}{\int \cos\theta \, d\Omega}$$
$$\approx \frac{\int_0^{\pi/2} \alpha L \sin\theta \, d\theta}{\int_0^{\pi/2} \cos\theta \sin\theta \, d\theta} = 2\alpha L.$$

Therefore the absorption path enhancement is
$$f_L = \frac{A_{single-pass}}{\alpha L} = 2.$$

The absorption enhancement factor is $f_A = f_I f_L = 4n^2$, the Yablonovich result. Note that $f_A = 4n^2 = C_2 n^2$ is consistent with (5).

For weakly absorbing material, absorptance is now given by
$$P = A = \frac{2 f_L \alpha L}{2 f_L \alpha L + \beta^{-1}} = \frac{4n^2 \alpha L}{4n^2 \alpha L + 1}. \quad (9)$$

The formula implies that the sunlight entering a $200 \mu m$ thick silicon film will have effectively 1cm optical thickness for absorption! Even for very weakly absorbing light, $P \rightarrow 1$; *such is the power of a single roughed surface*. However, for improved electrical properties of the solar cells, a much thinner absorber layer is desired—for such cases, even higher absorption enhancement is required to reach $P \rightarrow 1$.

We have used a configuration with planar, refracting surface facing the sun, and the roughened mirror at the back, because the ray tracing is intuitive and the analysis easy to explain. The results are unchanged if the configuration is reversed – a roughened refracting surface on top and a planar, reflecting surface at the back.

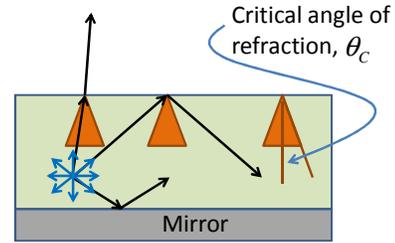

**Figure 6: Trapping of recycled photons.**

### F. Planar Surfaces and Photon recycling

The angle diagram (Fig. 5(d)) suggests that it is not necessary to have a random surface to achieve high photon intensity. Any process that scatters the photons away from the escape cone can achieve similar amplification. For example, if a photon is absorbed and immediately reemitted at a random angle, Fig. 6 shows that the number of repeated bounces will be identical to those from scattering by rough surfaces [6]. The randomization of the angles by a process called photon-recycling has been used with great success in creating ultra high efficiency cells that do not require rough surfaces [7].

## III. EXCEEDING THE $4n^2$ LIMIT

### A. Intensity Enhancement

One can increase $f_A (\equiv f_I \times f_L) > 4n^2$ by increasing $f_I$ as follows. The idea of this approach is to reduce the escape angle $\theta_{esc}$ below $\theta_C$, so that $P_{esc}$ is reduced, and the number of photons within the box is enhanced.

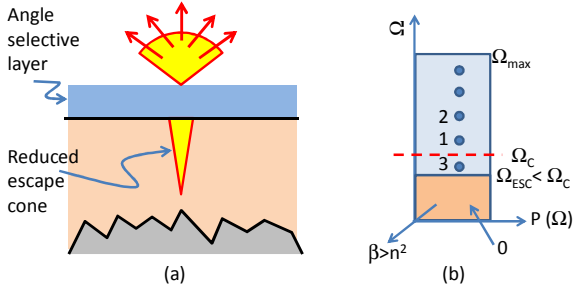

**Figure 7:** (a) Conventional structure with Lambertian back reflector yielding $4n^2$ absorption enhancement. An extra angle selective transmitter/reflector layer on the structure of can yield $>4n^2$ absorption enhancement as well as reduction in angle entropy loss. (b) The angle statistics shows that the suppressed escape angle allows for more states for photons inside the dielectric.

If the angle of the output emission angle is reduced by a factor $N_{out}$ by an angle selective layer so that light can be emitted with an angle of $(\theta_{Air} = 2\pi/N_{out})$, as in Fig. 7(a), the escape cone inside the dielectric will likewise be reduced by a factor $N$ ($\theta_{esc} = \theta_C/N$). Thus from Snell's law,

$$\sin\left(\frac{\pi/2}{N_{out}}\right) = n\sin\left(\frac{\theta_C}{N}\right). \quad (10)$$

We can simplify this relationship. For practical dielectrics we can approximate the critical angle as $(\sin\theta_C \approx \theta_C) = (1/n)$. Now if $N_{out}$ is large enough, we can re-write (10) as,

$$\left(\frac{\pi/2}{N_{out}}\right) \approx n\left(\frac{\theta_C}{N}\right) \approx \frac{1}{\theta_C}\left(\frac{\theta_C}{N}\right) = \frac{1}{N}.$$

Therefore,

$$N \approx \frac{N_{out}}{\pi/2}.$$

for $N_{out} \gg 1$. The roughened back reflector continues to randomize the light angles inside the dielectric, so that the angle-space in Fig. 7(b) is populated with equal probability. Following the derivation of Eq. (8), but now for the restricted escape angle, we find,

$$P_{esc} = \sin^2(\theta_{esc})$$
$$= \sin^2(\theta_C/N)$$
$$\approx (\theta_C/N)^2 = \frac{1}{N^2}\theta_C^2$$
$$= \frac{(\pi^2/4)}{N_{out}^2}\frac{1}{n^2}.$$

Recall that for a random surface with $D_S = 2$, $f_L = 2$. The absorption enhacement:

$$f_A = 2\beta f_L = \frac{2f_L}{P_{esc}} = N^2 \times 4n^2 = \frac{8N_{out}^2}{\pi^2} \times 4n^2.$$

This expression is in the form of Eq. (5), where $(\theta_C/\theta_{esc}) = N$ with $D_S = 2$ and $C_2 = 4$. The result is also similar to the absorption limit $(4n^2/\sin^2\theta)$ for concentrator solar cells [8] or cells with restricted emission [9–11]. Although the absorption path length enhancement is the same as before $(f_L = 2)$, the intensity enhancement is very high in this case $\left(f_I = (8N_{out}^2/\pi^2) \times 2n^2\right)$. If we could restrict the emission angle from $2\pi$ to the incoming solid angle of the sunlight $(\sim 6 \times 10^{-5}\,rad)$, then $N_{OUT} \sim (2\pi/\theta_D) = 10^5$, and

$$f_A = N^2 \times 4n^2 \approx (4 \times 10^9) \times 4n^2,$$

which is significantly higher than the Yablonovich limit. In this case, the photons are virtually guaranteed to be absorbed with probability 1 ($P \to 1$), because even the weakest absorbing materials have absorption coefficient of $\sim 10^{-5}/m$.

Returning to Eq. (3), we find that suppressing $\theta_{esc}$ not only improves $P$ (reduces the third term on right), but simultaneous suppresses the angle anisotropy and increase the open circuit voltage close to the bandgap [2], [10]. In practice, $N_{out} \to 10^5$ may be both impractical and unnecessary for absorption enhancement. The quadratic improvement of absorption with angle restriction ensures that even for moderate angle restriction, the absorbance increase is significant.

Note that Eq. (5) only holds when the rays are scattered such that all possible photon densities of state are accessible. For the following cases, this condition is not fulfilled and hence the absorption enhancement cannot be described by Eq. (5).

*B. 'Intensity' enhancement versus 'absorption' enhancement.*

The second approach to obtain $f_A(\equiv f_I \times f_L) > 4n^2$ is to increase $f_L \gg 2$ by preferential low-angle scattering of the rays. For these cases, $f_L$ is often reduced below $2n^2$, but $f_A(\equiv f_I \times f_L) > 4n^2$. We will now discuss the theory and two specific implementation that have been discussed in the literature [12], [13].

*1) Theory*

In all the preceding discussions, we have assumed the light to be scattered isotropically inside the dielectric. The intensity enhancement was $f_I = 2\beta = 2n^2$ for 2D random scattering surface. Average absorption path enhancement yields another factor of $f_L = 2$. Note that, the effective path length for light is much higher if it is scattered into large angles—allowing the rays to get absorbed. As shown in Fig. 8(a), rays with a smaller angle will require more number of bounces before absorption, having higher probability of escaping. A ray with an angle $\theta$ will undergo absorption of $(1-e^{-\alpha L/\cos\theta})$ for single pass through the dielectric. Now, assume a surface scatters the rays according to a probability distribution of $P(\Omega)$. Therefore, the average absorption path length enhancement is:

$$f_L = \frac{1}{\alpha L}\int (1-e^{-2\alpha L/\cos\theta})P(\Omega)d\Omega. \quad (11)$$

If the surface is designed such that it preferentially scatters parallel $(\theta \to \pi/2)$ to the surface, the absorption path enhancement $f_L$ can be very high. The strong absorption implies fewer bounces $(\beta < n^2)$, however, the overall absorption exceeds the Yablonovitch limit $(f_A = 2\beta f_L > 4n^2)$.

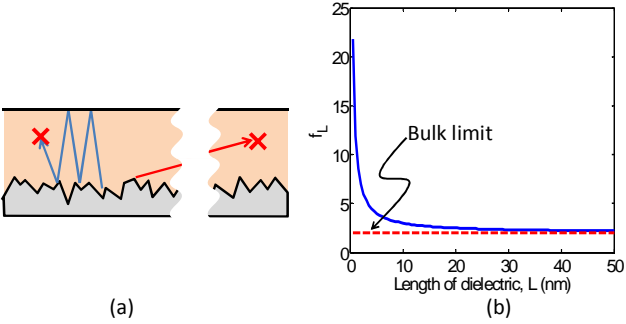

(a)  (b)

**Figure 8:** (a) Absorption of scattered light in the dielectric. (b) Absorption enhancement as a function of dielectric layer thickness *L*.

*2) Beating the Limit by Anisotropic Scattering.*

Several recent works beat the $4n^2$-limit of light absorption by innovative optical structures that scatters light predominantly into guided modes [12], as in Fig. 9(a). To understand this approach intuitively, we calculate the integral of $f_L$ shown above (11) by partitioning the rays into two groups: one for 'very low angle' evanescent wave absorption $(A_{ev})$, the rest for bulk absorption $(A_{bulk})$. Taken together:

$$f_L = \frac{A_{ev}+A_{bulk}}{\alpha L} \equiv f_{L(ev)} + f_{L(bulk)}. \quad (12)$$

Here, $f_{L(ev)}$ and $f_{L(bulk)}$ are path length enhancement due to evanescent mode and bulk absorption, respectively. We have already shown that for a 2D Lambertian surface $f_{L(bulk)} \approx 2$. The $A_{ev}$ does not depend on dielectric thickness, therefore, $f_{L(ev)} \sim 1/L$. For a moderately low absorption coefficient of $\alpha = 10^3 m^{-1}$, and low evanescent mode absorption $A_{ev} = 10^{-5}$, Fig. 8(b) shows that we can obtain very high $f_L$ as $L \to 0$, highlighting the importance of evanescent mode absorption.

The intensity enhancement $f_I$ is calculated by turning off absorption. Now the ray of light goes into the dielectric (arrow '0' in Fig. 9(b)), bounces at the back (states shown by set of dots marked '1') and then goes out (arrow '2'). Therefore, $f_I = 2\beta = 2$. In summary, while $f_I$ does not increase, $f_L$ does, so that overall $f_A$ exceeds the Yablovich limit.

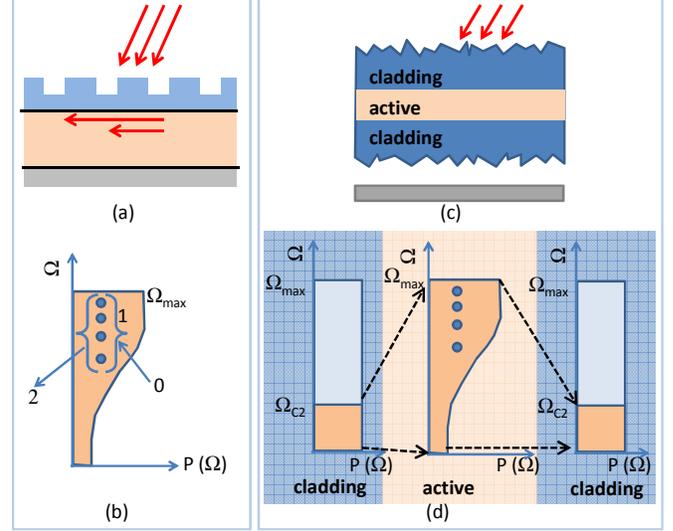

**Figure 9:** (a) Coupling of light into evanescent modes to reduce the probability of photon escape. (c) Thin active layer surrounded by high refractive index cladding for enhanced evanescent mode absorption. (b) and (d) show the angle statistics for (a) and (c) respectively.

Another form of such slot waveguide structure has been proposed by Green [13], see Fig. 9(c). The cladding layers with higher refractive index increases the evanescent mode coupling to the active layer with lower refractive index (see Fig. 9(c)), and increase $A_{ev}$. As the active layer is made thinner $L \to 0$, $A_{ev} \gg A_{bulk}$. As in Ref. [12], the increase in $f_L$ compensates for the decrease in $f_I$ to beat the Yablovich limit. The rays enter into the active layer form the 2 cladding layers through the escape cones of the cladding-active layer interface as shown in the angle statistics diagram of Fig. 9(d). The escaped rays into the active layer are distributed such that evanescent modes are increased. We note in passing that evanescent mode coupling in Ref. [13] is purely a wave optics phenomenon and there $A_{ev}$ can only be calculated by solving Maxwell's equations.

Note that, the absolute value of absorptance of these arrangements may not be high (i.e., not close to unity), although the absorption enhancement appears to be even orders of magnitude larger than the $4n^2$ limit.

## IV. SUMMARY

For weakly absorbing layer with a back mirror, the intensity enhancement limit is found to be in the form $f_A = C_{DS} \times n^{D_S} \times (\theta_C/\theta_{esc})^{D_S}$, where $n$, $C_{DS}$ and $\theta_{esc}$ are the refractive index, proportionality constant and the maximum escape angle, respectively, as defined by the dimensionality $D_S$ of the scattering surface. The derivation provides an intuitively simple explanation of the Yablonovich limit of $f_A = 4n^2$, as an special case with random refracting surface. Additional gain beyond this limit is possible if we observe the following: The essence of light trapping and intensity enhancement is related to the reduction of escape probability of the photons. This can be achieved either by increasing the number of states occupied by the photons inside the dielectric, or by decreasing the number of available states that allow photon escape. The example discussed in Sec. III(A) suggest that angle restriction provides significant additional gain because it not only improves absorption, but also the open circuit voltage/efficiency of a solar cell. It is important to remember that the additional gain is achieved only for normal incidence of sunlight obtained by orienting the cell towards the sun throughout the day

ACKNOWLEDGMENT

This material is based upon work supported as part of the Center for Re-Defining Photovoltaic Efficiency Through Molecule Scale Control, an Energy Frontier Research Center funded by the U.S. Department of Energy, Office of Science, Office of Basic Energy Sciences under Award Number DE-SC0001085.The computational resources for this work were provided by the Network of Computational Nanotechnology under NSF Award EEC-0228390.